\documentclass[a4paper, 10pt, conference]{ieeeconf}
\IEEEoverridecommandlockouts
\usepackage{biblatex}
\usepackage{graphicx}
\usepackage{dirtytalk}
\usepackage[utf8]{inputenc}
\usepackage[table,xcdraw]{xcolor}

\title{\LARGE \bf Distributed File System for an Edge-Based Environment }

\author{Rafael Neujahr Copstein$^1$ \and Fernando Luis Dotti$^1$
\thanks{$^{1}$School of Technology, Potifical Catholic University of Rio Grande do Sul (PUCRS), Porto Alegre, RS - Brazil.}%
}

\begin{document}

\maketitle
\thispagestyle{plain}
\pagestyle{plain}

\begin{abstract}

    Recent developments in the industry of personal computing led to a greater number of the so-called \textit{edge} devices. Such devices typically do not collaborate or foresee the possibility of collaboration to offer aggregated storage and computing capabilities. 
    The concept of distributed file system (DFS) is not new to the field of distributed systems, in fact, it is widely used in dedicated infrastructures, for example, in cloud computing applications. 
    
    In this work, we discuss reasonable assumptions for an environment composed of \textit{edge} devices, the main design issues and implementation challenges of a DFS in the given environment and how they would impact this application.   Thereafter we define a system model for an environment composed of \textit{edge} devices while taking into consideration their high mobility and common cases of network partitioning. Next, we describe an architecture for a DFS that withstands the proposed system model while offering most capabilities that a DFS at a dedicated infrastructure would.
    
    We conclude that the development of a distributed file system is a very complex task and, given the broad assumptions of the system model, 
    also hard to verify.  Some important aspects of the development lie as future work, but we believe that the developed DFS can be used not only as a tool on it's own, but also as a reference for further development of distributed file systems and, specially, of systems for infrastructures composed of \textit{edge} devices.

\end{abstract}

\section{Motivation}

    Recent developments in the industry of personal computing led to a greater number of client devices with computing capabilities, the so-called \textit{edge} devices \cite{edge}. Because of this, we can witness users owning an ever growing number of devices to fulfill different but often overlapping functions, and households containing an even greater number of them. These devices are ubiquitously equipped with networking functions and often with fully fledged operating systems. But, despite the existing knowledge in aspects such as networked and distributed systems, cooperation paradigms, heterogeneous platforms and other related topics, we observe that these devices typically do not collaborate or foresee the possibility of collaboration to offer aggregated storage and computing capabilities, but tend to focus on their specific functions. Besides that, the specific functions of these devices are often very limited in spite of the fact that there is usually a considerable amount of computing power.
    
    Whatever the reasons behind this state of affairs may be: economic driven (devices are cheaper and cheaper), market driven (industry wants to stick to closed families of devices), or even the opinion that collaboration at edge devices is difficult to handle due to disconnections and more ephemeral life cycles; it is not difficult to agree that more elaborated collaboration models among such devices are lacking and are needed in order to achieve higher functions, which are likely impossible to be offered by isolated devices.
    
    The idea of using the combined capabilities of devices is not new, in fact, dedicated infrastructures for cloud computing use the combination of many mid-ranged devices to offer processing and storage resources that exceed those of an off-the-shelf device. We call an infrastructure \textit{dedicated} when it is being actively maintained by a company, it is fault-tolerant (inherently using replication), highly available, and the primary purpose of the devices that compose it is to collaborate in solving a task (as opposed to being personal devices or a workstation). In this kind of infrastructure, particularly for combining storage capacity from multiple devices for data modeled as files and folders, we have what is called a \textit{Distributed File System}.
    
    Distributed File Systems (DFS), according to Sinha \cite{sinha}, offer an abstraction similar to that of a conventional file system for a distributed environment where devices are inherently dispersed. It supports remote information sharing (where a file can be accessed transparently by any node in the system), user mobility (the user has the flexibility to work from any node at any time), availability (the system should be available despite the eventual failure of a node -- which usually falls back to the infrastructure the system is installed in), and diskless workstations (the system can be accessed by machines with no long-term memory). Besides that, a good DFS should also be transparent in terms of structure (nodes can be added and removed from the system without impacting its use), performance (the system has to be at least as fast as a conventional file system), and location (the physical location of nodes shouldn't impact the system's performance).
    
    Considering the concept of DFS and the need for a collaboration model for \textit{edge} devices, it is reasonable to ask: (i) What are reasonable assumptions for an environment composed of \textit{edge} devices and how do they impact a DFS? (ii) Which kind of file system semantics should and could be supported? (iii) What are the main design issues and implementation challenges for a DFS in such environment?

\section{Related Works}

    \begin{table*}[t]
    \begin{center}
    \caption{Comparison of Available Distributed File Systems}
    \label{related-works-taxonomy}
    \begin{tabular}{|l|c|c|c|}
    \hline
    \multicolumn{1}{|c|}{\textbf{File System}} & \textbf{File Consistency} & \textbf{Geographical Scalability} & \textbf{Centralized} \\ \hline
    Google File System \cite{googlefs}                        & Strong                    & WoL                               & YES                     \\ \hline
    GlobalFS \cite{globalfs}                                   & Strong                    & SoW                               & NO                      \\ \hline
    XTreemFS \cite{xtreemfs}                                   & Strong                    & WoW                               & NO                      \\ \hline
    QuantcastFS \cite{quantcastFS}                                & Eventual                  & WoL                               & YES                       \\ \hline
    AFS \cite{afs}                                        & Weak                      & WoW                               & NO                       \\ \hline
    PVFS \cite{pvfs}                                       & Read After Write          & WoL                               & NO                       \\ \hline
    
    \end{tabular}
    \end{center}
    \end{table*}

    The concept of a distributed file system is not new, in fact, there are many examples of such in the literature. Distributed file systems tend to vary in terms of geographical scalability, data consistency, architecture, and others. In this section we explain some of these concepts and do a brief overview of some publicly available distributed file systems.

    In the context of distributed systems, we say that a system is \textit{geographically scalable} when it is able to run -- and perform relatively well -- even when the nodes that compose the system are physically distant from one another. In Pacheco \textit{et al.} \cite{globalfs}, geographical scalability is measured as \textit{geographical scaling potential} and is divided in three levels: systems that work on LANs (\textit{WoL}), systems that support but perform poorly in wide area networks (\textit{WoW}), and systems that scale in WAN (\textit{SoW}). In this work, we will use this classification to refer to geographical scalability.
    
    Since files are accessed from multiple nodes, distributed file systems use the concept of \textit{consistency} to establish semantics when there are modifications to a file. As stated in \cite{consistency}, we say a system has \textit{strong consistency} when all the machines are only able to access -- and modify -- the latest version of a file, in other words, the file is always up-to-date from every perspective. \textit{Weak consistency} means that different nodes will process file modifications in (possibly) different orders, that is, different nodes might see different versions of a file at a given time. 
    
    According to Tanenbaum and Van Steen \cite{tanenbaum}, when a system offers \textit{eventual consistency} it means that, if there are no further changes to a file, eventually all nodes will see the updates done by every other node. Finally, \textit{read-after-write} consistency (or \textit{read-your-writes}, according to \cite{tanenbaum}) makes sure that data written by a process, $p$, will be immediately available to any successive read operation by $p$.
    
    These models for file consistency are used when the systems supports replication, that is, there are multiple copies of the same file stored in different machines. Along with file consistency for replication, a DFS must, even if it does not support replication, establish semantics for file sharing, since shared access to files is inherent to the definition of DFS. File sharing semantics can be, according to \cite{tanenbaum}: UNIX semantics, where every operation is instantly visible by every process; Session semantics, where no operation is visible to other processes until the file is closed; Immutable Files, where no modification is possible; and Transactions, where all changes occur atomically.
    
    An important aspect of the architecture of a distributed file system is whether it is a centralized architecture or a decentralized one \cite{tanenbaum}. Having a centralized architecture determines whether the system has a master node -- that oversees most operations and coordinates the other nodes -- or if the system has equally powered nodes that work together.
    
    As seen in table \ref{related-works-taxonomy}, there are distributed file systems with different characteristics available in the literature. Despite being common sense that a distributed file system with strong file consistency and \textit{SoW} scalability is more desirable due to an easier to comprehend operation, it can be seen that achieving both is very difficult. Out of the given examples, only GlobalFS \cite{globalfs} was able to provide both.

\section{System Model}

    An environment composed of \textit{edge} devices is different from the typical environment for distributed computing in the sense that the devices that compose it are constantly moving and changing their state with regards to their connectivity. Due to this, some assumptions are made regarding this kind of environment such that they must be respected for a consistent execution of any system. From this point forward, we will refer to an environment composed of \textit{edge} devices as an \textit{edge}-based environment.
    
    First, we assume an environment with no global time and no clock synchronization. Message delivery can be affected by unknown delays and message loss is also possible, though fair-loss is assumed. Lack of response is not considered a failure in this work, but it is interpreted as temporary unreachability. Unreachability, here, is the state where two machines cannot communicate due to the lack of routing paths between the two across the network.
    
        
    
    Second, we assume that \textit{edge}-based environments must frequently deal with a \textbf{high churn rate}. In other words, devices that compose this environment are constantly moving in and out of reach of each other from a network connectivity perspective. Here, the term \textit{churn} does not apply to birth/death of processes, but to their ever changing reachability status (though birth/death rate could become high as well). We also assume that this environment will inevitably experience episodes of \textbf{network partitioning}. We say that a partition occurs when distinct groups of devices become completely unreachable from one another, that is, no device from one group is able to reach any device from the other.
    
    
    
    
    We also assume the \textbf{lack of centralization}, that is, a system that runs on an \textit{edge}-based environment can't assume the existence of a highly reliable device to oversee most operations that happen during the execution. This would mean that such a device would have to be reachable from every other device every time an operation were to be done. Due to the high mobility of \textit{edge} devices, this would be counter-intuitive.

    Since there are many scenarios where some devices are not able to connect to one another, providing strong consistency would, likely, be a very complex task and could impact the system's operation negatively. We do, however, assume that any correct device (say one that won't leave the system forever) will, eventually, reconnect with a sufficient number of devices for a sufficient amount of time such that it is available for making progress. Thus, a system installed on an \textit{edge}-based environment should be able to offer eventual consistency.

\section{Proposed Solution}

    Considering the problem of providing distributed file system (DFS) capabilities in an infrastructure composed of \textit{edge} devices, we developed a DFS specifically for this kind of environment. Our solution was built to withstand the characteristics presented by the system model as follows:
    
    \subsection{High Churn Rate}
    
        The division of the task of storing files is inherent to the very concept of DFS. In this sense, when two devices move out-of-reach of one another they can no longer rely on the possibility of reaching that device's share of the task, that is, if there are files stored in one device, then those files become unavailable for devices that have gone out-of-reach.
        
        Considering this, the developed DFS utilizes a module for detecting \textit{device reachability}, that is, a module that continuously updates itself to provide information on which devices are reachable and which are not. When a device becomes unreachable, the DFS hides all of the files that are stored in that device since any kind of operation over those files wouldn't get completed.
        
    \subsection{Lack of Centralization}
    
        Since assuming the existence of a central device to coordinate the system is less than ideal, the devices that compose the DFS are responsible for spreading the information regarding which devices are a part of the system and which files are available on the system, that is, the system's metadata. 
        
        The devices that compose the system are named \textit{members}, and each member stores references to a set of other members it knows are a part of the system. Whenever a new device enters the system it gets introduced to the other members by receiving another member's list. Whenever a member's list changes, that member is responsible for propagating the changes to the other members it knows.
        
        In terms of the of files available in the system, they are called the DFS's \textit{logical hierarchy}. Again, each member is responsible for maintaining a registry of the current state of the hierarchy. However, the changes are only propagated if they affect the files \textit{owned} by that member, that is, they are physically stored on that device.
            
    \subsection{File Ownership}

        The developed DFS, despite being operable from every device, has to have it's files physically stored in one or more devices (in the developed DFS, it is always one). We say that the member that stores a file is that file's \textbf{owner}.
        
        A file's owner is responsible for managing that file's metadata and sharing any modifications with the other members. Say a user wants to rename a file, then that intention has to be sent to the owner so that it can change the file's name and correctly propagate that change. Other members cannot execute this operation without the owner's consent.

        When two members become unreachable from one another, the files they own become unavailable to the other member. This happens so that the user cannot attempt to operate on a file whose owner is unreachable and, thus, cannot consent to any operations. Folders are only an organizational mechanism, so they are not owned by any member.
 
    \subsection{Network Partitioning}
    
        With the previous mechanisms in place, we can assume that users can only operate with files that are, indeed, accessible to them. So, if a member loses connectivity for whatever reason (it gets lost, stolen, crashes, etc) the other members will update to a consistent state (where unreachable files are not available to the users). If there is a partition, then members will see different hierarchies, since different portions of it are available to them. 
        
        However, once previously disconnected members reconnect (the partition is undone), the DFS becomes responsible for updating the logical hierarchy to a correct state (where every reachable file is also available). When this kind of situation occurs, we say that members are synchronizing the state of their hierarchy. 
        
        The synchronization of two hierarchies may result in conflicts, that is, two files at the same path owned by different members. In that case, the system keeps the entries for both files in the hierarchy but resolves their names to something unique to avoid confusion. Once the conflict has been solved (one file has been renamed, moved or removed) the original name of the file is restored.
        
    \subsection{Consistency}
    
        Regarding consistency, we say that the developed DFS is \textit{eventually} metadata consistent. As assumed by the system model, correct devices will eventually reconnect for making progress. When that happens, the system will synchronize the multiple versions of the logical hierarchy and of the members list among all connected devices resulting in the same metadata on all members.
    
        In terms of data (the files themselves) the system's consistency depends heavily on it's implementation. The current implementation of the developed DFS uses NFS (which is detailed in a later section) for remote file access, which offers \textit{weak} consistency, that is, multiple users can see different versions of the same file at the same time. The last write to a file is the one that remains, even if some previous writes were not received by the process that wrote last. However, if the module for remote file access were to be re-implemented as something that supports eventual consistency, for example, than the DFS would be eventually consistent regarding files.
        
    \subsection{Security}
    
        The developed DFS assumes that the system will be installed on trusted devices that communicate over a trusted network and does not foresee any kind of byzantine behaviour. Considering that the members of the system must be able to send and receive information to one another, we assume they are all routable, that is, that they possess a routable IP address to reach them. With that, we can assume that the developed system will, most likely, be installed for use in local networks, which are usually trustable. 
        
        If that is not the case, the system could be secured to some extent by creating encrypted tunnels between the members for communication. This would prevent, to some extent, that attackers create messages in the name of another member of even that they eavesdrop on the process of sending and receiving file content.
        
        A more robust security model for the developed DFS lies as future work.
    
    \subsection{Replication}
    
        In the developed DFS (and in distributed systems in general), maintaining a consistent state across multiple processes is challenging. Adding to the fact that, in this system model, processes may be unreachable and that modifications could be made from any device at any time, we end up with an intricate set of constraints. The addition of  support for replication, inherent to fault-tolerance, is directly affected by such constraints.
        
        Due to a limited amount of time for the development of this work and given the scope of the problem, support for replication of files was not included. Despite that, replication would come as a very important addition to the improvement of the system's reliability and, therefore, lies as quite important future work.
        
\section{Architecture}
\label{sec-architecture}

    \begin{figure}[b]
        \centering
        \includegraphics[width=0.35\textwidth]{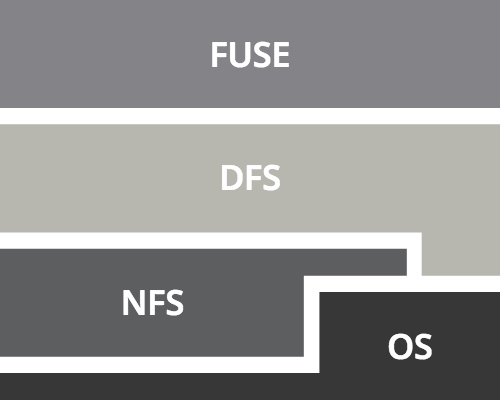}
        \caption{Overview of the system architecture}
        \label{fig-architecture}
    \end{figure}

    The architecture of the developed DFS is composed of three main layers: the interface layer, the logical layer, and the file management layer. Due to time and resource constraints, as well as aiming at an easy integration with existing standards,  we decided to use pre-existing tools for the interface and file management layers, in this case, respectively, FUSE and NFS. Both these tools provided just enough functionality so that a viable prototype could be developed. Despite having to, sometimes, adapt to how these tools work, the layered nature of this architecture allows for changing the implementation of any layer with relative ease. An overview of how the layers are structured in the architecture can be seen in figure \ref{fig-architecture}.
    
    \subsection{FUSE}
    
        The \textbf{Filesystems in Userspace} (FUSE) \cite{fuse} library allows for the creation of file systems without having to deal with kernel code on linux-based systems. One of the primary uses of FUSE is to provide different kinds of information under the paradigm of files and folders. For example, a system could show different reachable networks as folders and the hosts connected to each one as files within. Operations like moving a file to a different folder could mean that a host is supposed to get disconnected from that network and be connected to a new one. 
        
        FUSE proved to be very useful in the development of the DFS because it requires only that some functions are implemented for dealing with common file system operations like read, write, get attribute, etc. In our case, the implementation of these functions would retrieve the entry for a given file from the metadata and resolve the actual path to where that file was stored and, only then, actually perform the requested operation. How to retrieve the file's entry and how to access the file if it is stored in another machine are responsibilities of the other layers.
        
        When executing, the fact that FUSE was used is mostly transparent, with the DFS being shown similarly to an external device (like a thumb drive, for example). Since it is seen as a device by the file system, it will get properly unmounted before shutting down, allowing for proper memory cleanup to take place.
    
    \subsection{NFS}
    
        The \textbf{Network File System} (NFS) \cite{nfs} is a tool that allows external hosts to access a given portion of a machine's file system. It allows the original machine to specify which hosts are able to see the files, which are able to modify (if any), and some other parameters regarding consistency and permissions. Files shared via NFS are seen as regular files on the external hosts' file systems, which allows applications to use such files with no regards to whether they are physically stored in the same device or not. This is another great advantage of NFS: every operation happens over the internet, without requiring the external host to copy the file to a local drive. Under the right conditions, dealing with files over a network is not noticeably less performant than dealing with local files, which allows machines with reduced storage capacity to access a larger amount of data.
        
        NFS proved useful in the development of the DFS because it allows for a simple way of accessing files on different hosts through the network. That means we didn't have to worry about developing protocols to send and receive chunks of data ourselves, that was already handled by NFS. Despite all these benefits, NFS does lack more robust mechanisms for identifying the hosts inside a network (avoiding doppelgängers) and for securing it's traffic (NFS packets are not encrypted). There are a couple of recommended solutions to these problems like using the Kerberos \cite{kerberos} protocol for host identification or using ssh tunnels for secure communication. The Kerberos protocol requires a centralized infrastructure, which is not the case of \textit{edge}-based environments, though the use of ssh tunnels could be explored as future work.
    
    \subsection{Logical Layer}
    
    
        The system's logical layer is, arguably, it's core. It's also, the system's bottleneck, that is, most of the system's performance is related to how this layer performs.
        
        This layer is responsible for handling the metadata, which, in turn, is what dictates the system's behavior. The metadata can be broken into two main areas: \textit{members} and \textit{hierarchy}.
        
        \subsubsection{\textbf{Members}}
        
            Because of the high churn rate inherent to the \textit{edge}-based environment, the list of members that compose the system is ever changing. Besides that, the system must also track the reachability of each member so that some action can be taken when any of them becomes unreachable (make the files it owns unavailable, for example).
            
            Each member is represented by an unique name, an IP address and a port number. The IP and port combined should point to the machine where that member is executing at a port that it is monitoring. This information is provided by the member when it starts executing.
            
        \subsubsection{\textbf{Joining the system}}
       
            In order to join the system, a new member must, first, request a name from a member that is already established in the system. Since this request can be made to any member, that member must guarantee that no other member has been given that same name before. This is solved by the \textit{name} protocol, explained in a later section. After being named, the two members add each other to their respective lists of members which, as the name suggests, is stored as a list in memory.
  
            The list of members, besides storing all the members known, is also identified by a sequence number. Whenever the list changes (a member gets added) the sequence number increases. This is used so that other members can know when there are (possibly) new members it does not know of in someone else's list. Members periodically poll each other for this number and, when they notice that it has increased, request that member's current list. This flow is handled by the \textit{sync} protocol.
            
        \subsubsection{\textbf{Detecting Reachability}}

            In the field of distributed systems it is common to use failure detection modules to assess the state of other members and handle any failure that may happen. In the developed DFS this is no different, though we refer to our failure detection module as a device reachability detection module, since we are not detecting failures per se, but detecting whether a member is within reach of another, which could be impacted by, but is not necessarily related to, an eventual failure. Here, messages are sent periodically to the other members. Receiving such a message means that the sender is active and within reach. If a message like this is not received from a certain member for some time, it is assumed that that member is out of reach and the proper actions are taken. These messages are sent even to members considered out of reach, since they could reconnect at any time as stated by the system model. This behavior is implemented by the \textit{ping} protocol.

        \subsubsection{\textbf{Hierarchy}}
        
            In regular file systems, the file/folder hierarchy is somewhat related to the way the data is stored physically. In the developed DFS files are stored in different machines independently from where they are located in the hierarchy. For this reason, each file in the hierarchy must have an individual entry in memory stating which member owns that file and it's physical name, which does not have to match the name in the hierarchy. Because the names don't have to match and the location in the hierarchy does not reflect the physical location of a file, the hierarchy is also referred to as \textit{logical hierarchy}.
            
            The logical hierarchy is stored in memory as a generic tree data structure where each branch node is a folder and each leaf is a file. That way, a file path can be used to traverse the tree towards that file's entry.
            
            
            
        \subsubsection{\textbf{Propagating Hierarchy Changes}}
            
\begin{figure}[t]
                \centering
                \includegraphics[width=0.25\textwidth]{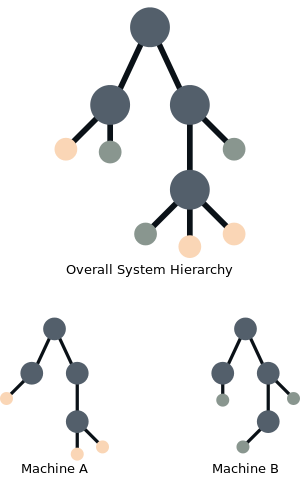}
                \caption{Multiple trees combine into a single logical hierarchy}
                \label{fig-hierarchy}
            \end{figure}

            The logical hierarchy follows a similar mechanism to the list of members for tracking changes -- it uses a sequence number -- however, each member manages a different number which only reflects the changes regarding the files owned by that member. Say a file is renamed, then the sequence number will only change for the member that owns that file. This stops members from propagating the same change multiple times and, possibly, causing sequence numbers to grow indefinitely. Once a member receives a larger number than what it knew to be a member's sequence number, it requests a synchronization with the sender. The sender, then, sends a description of it's version of the hierarchy tree. For the same reason as before, the tree that gets sent includes only files owned by the sender -- the other files/paths are not included in the message. The synchronization of hierarchy trees is handled by the \textit{sync} protocol. The combination of multiple hierarchy trees into the logical hierarchy is illustrated in figure \ref{fig-hierarchy}.
            
        \subsubsection{\textbf{File Operations}}
            
            Whenever the interface layer, in this case, FUSE, sends a call for the creation of a file, the member that is processing this call must determine where to store that file and how to place it in the logical hierarchy. The first step is to check for a name conflict: If there is a file placed in the same path as the one being created and both files share the same name, the call for file creation is cancelled and an error is returned. If there are no conflicts, the current member uses a member selection heuristic to determine which member will become that file's owner. The current heuristic randomly selects a member from the list of members and forwards the call to it. Since calls from the interface layer block the execution of the file system until they are responded, the current member must wait for a response from the assigned file owner before it can respond the interface layer's call. 
            
            The assigned file owner, once it receives the request for file creation, physically creates the file locally and adds it to the logical hierarchy from it's point of view. The information regarding this new file will later be propagated using the hierarchy change propagation mechanism explained previously. Once this is complete, the assigned owner responds with a message of success or an error code which will be responded to the local interface layer. Errors can include not enough memory, name conflicts, invalid path, and others.
            
            As stated in the developed DFS's model, any changes to a file must be handled by it's owner. For this reason, both deletion and renaming execute in a similar way to file creation, but instead of choosing a member according to a heuristic, since the files already have an owner, then that owner is sent the call. The protocol that handles these three operation (creation, deletion and renaming) is called \textit{freq}.
            
            Since the physical organization of files does not necessarily reflect that of the logical hierarchy, whenever a file is open it has got to have it's name translated from logical to physical. In other words, the logical name is used to traverse the hierarchy and to retrieve the file's owner and the associated physical name (used by the file management layer). Once that is done, a standard file system call is made for opening the physical file. Further operations, such as reading and writing, are done directly to the file handler, which bypasses the logical layer and are handled by the file management layer using standard operating system calls. 
            
        \subsubsection{\textbf{Handling Name Conflicts}}
            
            Since members can be out-of-reach while performing file operations, it might happen that they inadvertently create a state of conflict, that is, create equally named files at the same location. Since they cannot communicate with one another, they cannot warn each other about the conflict that is about to happen, so both operations are allowed. This creates a situation where, when members synchronize their hierarchies, they end up with files with the same name at the same location but with different owners (and different physical counterparts). 
            
            Suppose there are three disconnected members: $A$, $B$ and $C$. Say $A$ creates a file named \textit{Jupiter.jpg} at the root of the file system and connects momentarily with $B$ for enough time so that they synchronize. Next, $C$ creates a file with the same name also at the root of the file system and also connects for enough time to synchronize with $B$. Now, $B$ has two equally named files from different owners and must determine how to display them to the user. Most operational systems will not allow for two different files to be named the same at the same location and, even if they did, it would cause the user a lot of confusion.
            
            For these kinds of problems, the frequently proposed solution is to keep one of the files and to delete the other. However, due to network partitioning, a solution like this could end up causing members that are connected to only one owner to not see either file, because one of them got removed and the other is not accessible. In our example, if $A$'s file got chosen as the one to keep and $C$'s file got removed, then $C$ would not be able to access either file.
            
            The developed DFS solves this problem by flagging the files as being conflicted. This allows the interface layer to present both files in a way such that the user can recognize the conflict but, at the same time, operate with both files. Whenever this conflict is solved, either by renaming or deleting one of the files, the conflicted files get unflagged. Whenever the logical hierarchy gets manipulated, checks for conflicted files happen to make sure they are correctly flagged (or unflagged).

\section{Protocols}
\label{sec-protocols}

    As mentioned in the previous sections, protocols are used for communication between members and are the ones responsible for solving many issues that may appear. Each protocol is identified by a four-letter name that is written at the beginning of every message in the protocol so that it can be routed correctly within the application. In the following sections each protocol is presented along with an explanation on why it solves eventual problems.

    \subsection{Protocol \textit{name}}
    
        When a member wants to join the system it must, first, introduce itself to another member already established in the system. However, members are identified by their names, and a brand new member isn't named before joining the system. Even if a name got picked for the new member, there could be a problem where two new members are named the same -- which is undesirable. For this reason, a new member starts executing without a name and, when it introduces itself, it does so with a placeholder name (even empty) using the \textit{name} protocol.
        
        The \textit{name} protocol consists of a newcomer (a member that wants to join the system) and a host (a member that is already established in the system). First, the newcomer sends a message to the host containing it's IP address and a port number that it is listening at. The host, then, generates a unique name for the newcomer by appending a locally unique sequence of characters to the end of it's own name. The names are built in a hierarchical way (the host's name is contained in the newcomer's name) so that, since a member's name is unique, uniqueness can be assumed inductively. Once the name is generated, a message is responded to the newcomer with it's newly assigned name and the name of the host. After this exchange is complete, both members can add each other to their list of members, which triggers the other protocols.
        
    \subsection{Protocol \textit{ping}}
    
        Once two members have been properly introduced to one another, they begin to periodically send each other messages containing their names to confirm that they are still connected and within reach (this is part of the device reachability detection module). After a certain amount of these periods has passed (in the current implementation, 4) without a message being received from a member, that member is flagged as inactive. Whenever a message does get received, this count gets reset. This process repeats itself for the lifetime of the member.
        
    \subsection{Protocol \textit{sync} (Members)}
    
        After a member has been introduced to the system it must be updated with the current state of affairs. Besides that, all members other than the host must also be introduced to the newcomer. As stated before, this is done by providing a sequence number that represents the current version of a member's list of members. For performance reasons, this number is attached to the \textit{ping} protocol messages. Once a member (the receiver) receives a sequence number from another member (the provider) that is greater than the ones it has on record, it sends the provider a \textit{sync} request message.
        
        A \textit{sync} request message includes the receiver's name and a flag indicating that this message is a \say{request}. Once a request is received by the provider, it generates a message reporting the current state of it's list of members which, in turn, gets attached to a \textit{sync} reply message along with the provider's name and the current sequence number of the list of members. This message gets sent to the receiver. Once the message is received, the receiver compares the received list with it's current list and updates the sequence number on record for the provider. If the received list contains any members that are not contained in the current list, then the receiver adds those to it's own list and updates it's own sequence number. If there are no new members, than this number does not get update. This happens in order to avoid indefinite propagation of lists of members.
        
    \subsection{Protocol \textit{sync} (Hierarchy)}
    
        A very similar mechanism to the one used to synchronize lists of members is used to synchronize file hierarchies. For this reason, the synchronization of hierarchies also happens under the \textit{sync} protocol. The main difference from syncing lists of members is that, this time, the sequence number represents the current state of the logical hierarchy regarding only the files owned by that member. Since other members cannot modify files owned by another member without it's consent, a synchronization of the hierarchy with another member will never impact the sequence number.
        
        Another big difference between the two uses of the protocol is the that the syncing is more complex than simply adding the missing entries. It starts by reading the top of the received hierarchy and, for every folder, verifying if it also exists in the current hierarchy. If it does not, then it gets added at that location. The contents of a folder are synchronized all at once.
        
        Since the logical hierarchy is stored in memory as a linked structure, consider the contents of a folder as a linked list. Whenever a folder is going to get synced, there are two linked lists: the current list of files in the hierarchy and a list of the files owned by the provider. We start by pointing at the first file in each list such that it is owned by the provider. In the received list, it's going to be the first item. In the current list, it might be anywhere (if at all). If the two files share the same name, they are the same file, which means that the file in the current list should remain there. If that is the case, the protocol stores the sequence number for the synchronization along with the current file and moves on.
        
\begin{figure}[t]
            \centering
            \includegraphics[width=0.45\textwidth]{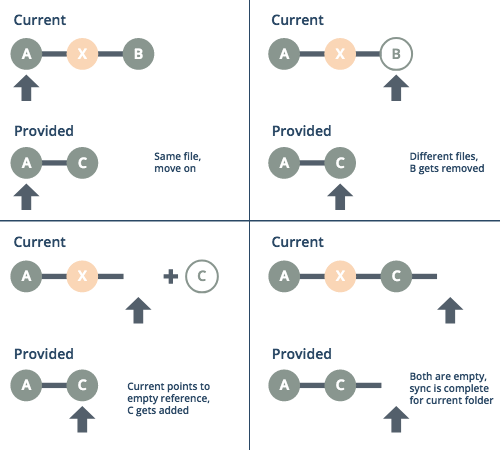}
            \caption{Overview folder sync process}
            \label{fig-sync-folders}
        \end{figure}

        If the files are different, then it can be assumed that the owner of the files is the source of truth, that is, it means that the current file can be removed. In that case, the provided list will not point to the next item in the following step, it will remain in place to assure that the file gets synchronized properly. If the current list points to an empty reference, that is, there are no more items owned by that member in the current list, then every file from that point forward in the provided list are new files, so they get appended to the end of the current list. If the provided list points to an empty reference while the current list does not, they are considered different, so the current one gets removed (and so will all of the following files owned by the provider in the current list). This process repeats until both lists are pointing to empty references. The process of syncing a folder is illustrated in figure \ref{fig-sync-folders}.
        
    \subsection{Protocol \textit{mont}}
    
        The \textit{mont} protocol exists exclusively because of the use of NFS. This protocol is used so that members can request access to a folder being shared by another member via NFS. The \textit{mont} request message includes the name of the requesting member, which should already be included in the requested member's list of members. If it is not, then the message is ignored. Once a request arrives from a known member, the requested member changes the proper NFS settings so that the requesting member's IP becomes able to access the requested member's files. Once that is done, a \textit{mont} reply message is sent to the requesting member informing that it now has access to the requested member's NFS shared files. The requesting member then proceeds to actually mounting the remote files, that is, it makes them available as if they were regular files in the underlying file system, even though they are stored remotely.
    
    \subsection{Protocol \textit{freq}}
    
        When a member wants to modify a file that it does not own or it wants a new file to be created but it is decided that some other member will own it, it has to contact the owner (or the assigned owner) of that file so that the operation can proceed. The \textit{freq} protocol (which stands for \textit{file requests}) handles the communication between members when one of them wants to perform one of three kinds of operation: creation of a new file, deletion of a file, or renaming of a file. The request message for this protocol includes the path of the file to be operated on and the type of operation (add, delete or rename) at a minimum. If the operation is a rename, the new name is also included.
        
        Unlike the other protocols, the \textit{freq} protocol is the only \textit{blocking} protocol, that is, the communication is done over a reliable connection that blocks the process until it receives a response. That means that whenever one of these operations is taking place, no other operation is happening on the member that requested it. This is so because the interface layer, FUSE, expects a return value to any call it makes reporting either success or failure of the operation and, depending on the result, this affects what the user sees. Since the protocol waits for the operation to be completed on the file's owner, either as a success or as a failure, it is able to provide such information to FUSE which, in turn, provides a coherent user experience.
        
        Any of the possible operations, if they succeed, will result in a change to the requested member's hierarchy. Those changes will be properly propagated using the aforementioned \textit{sync} mechanism.
    
    \subsection{Handling the system model}
    
        Considering the system model, there may be scenarios where some members experience network partitioning or message loss while executing some of the aforementioned protocols. Even though the execution may get interrupted by such events or some messages may get sent more than once, the developed DFS guarantees that every member will remain in a consistent state, as we argue below.
        
        Regarding synchronization, the protocol \textit{sync} only executes when the complete information is available, that is, once the logical hierarchy from the other member has been entirely received. If that is not the case, say because of partitioning happening during the information exchange, then the protocol's execution gets interrupted and the logical hierarchy does not get affected. If a \textit{sync} request is sent multiple times, say due to network delays causing the member to retry, then there will be no side effects other than multiple replies being sent by the target member. If multiple \textit{sync} replies are received, then the protocol relies on the information's sequence number to avoid synchronizing old information into it's metadata. Replies with a sequence number smaller than the current one are not considered for synchronization. Since they deal with different sets of files, replies from different members can be received, and synced, in any order.
        
        For file requests, the protocol \textit{freq} executes in a blocking manner, that is, if there is a partition that makes members executing this protocol unreachable, then they would be blocked until a reconnection happened. To avoid such situations, \textit{freq} relies on timeout mechanisms provided by the underlying network protocol, in the case of the developed DFS, TCP. 
        
        Given that members are in a consistent state before the execution of the protocol, the \textit{freq} protocol guarantees that, even if the members become unreachable and the protocol halts prematurely, their state will always end up being consistent. This is done by executing each step that modifies the metadata locally, that is, in a way that no communication with the other members is required. This guarantees that, even if no message can be sent or received, the steps are completely executed (we are not considering device crash here).
        
        If a file request gets sent multiple times, the very nature of the operations will make it idempotent, that is, creating a file that is already created, deleting a file that has already been deleted and renaming a file to a name it already has are all naturally idempotent operations. If a delay large enough is assumed so that a message from another member is received in between these multiple messages, nothing can be assumed regarding idempotence since the state could have changed before one of the multiple messages is processed. Even though the state in unpredictable, since messages are being fully received, the state will not end up inconsistent.
    
\section{Conclusion}

    The development of a distributed system increases in complexity relatively to complexity of the system model. In an \textit{edge}-based environment, where there are many different possible scenarios, developing a distributed system that remains in a coherent state during its execution can be a very complex task. When the distributed system operates on such a fundamental aspect of modern computing, file systems, on top of all the difficulties brought by the system model there are many pre-conceived expectations on how the system should behave and perform.
    
    In the developed DFS we have proposed solutions to many of the problems that appear in the presented system model. The development of a prototype also helped demonstrate the correctness of most of these solutions and helped us adjust some of them during the conception phase of this project. The developed system shows some limitations due to the use of pre-existing tools to solve some of the functions of the system and due to the many different scenarios that have to be considered in the given system model. Overall, given the limited time span available for the development of this project, the tradeoff is acceptable.
    
    Some important tasks lie as future work such as a proper security model and support for replication, required for eventually making the developed DFS fault-tolerant. Despite that, we believe that the developed DFS can be used not only as tool on it's own, but as a reference for further development of distributed file systems and, specially, of systems for \textit{edge}-based environments.

\section{Acknowledgements}

    We would like to thank professor \textbf{Sérgio Johann Filho} for his help in the development of this work. Because of his suggestions regarding the system's architecture, the Linux kernel and the use of FUSE, we were able to develop a prototype that is compatible with modern UNIX-based operating systems.
   
    We would also like to thank the help of \textbf{Brian Yip} in the development of the algorithm for resolving file conflicts and the help of \textbf{André Antonitsch} while validating the model against it's many possible scenarios.

\addtolength{\textheight}{-12cm}  

\printbibliography

\end{document}